\documentclass[12pt]{article}
\usepackage{epsfig}
\usepackage{amsmath}
\begin{document}
\begin{center}
\LARGE
\textbf{Understanding Long-Distance\\
Quantum Correlations}\\[1cm]
\large
\textbf{Louis Marchildon}\\[0.5cm]
\normalsize
D\'{e}partement de physique,
Universit\'{e} du Qu\'{e}bec,\\
Trois-Rivi\`{e}res, Qc.\ Canada G9A 5H7\\
email: marchild$\hspace{0.3em}a\hspace{-0.8em}
\bigcirc$uqtr.ca\\
\end{center}
\medskip
\begin{abstract}
The interpretation of quantum mechanics
(or, for that matter, of any physical theory)
consists in answering the question:
How can the world be for the theory to be true?
That question is especially pressing in the case
of the long-distance correlations predicted
by Einstein, Podolsky and Rosen, and rather
convincingly established during the past decades
in various laboratories.  I will review four
different approaches to the understanding of
long-distance quantum correlations:
(i) the Copenhagen interpretation and some of
its modern variants; (ii) Bohmian mechanics of
spin-carrying particles; (iii) Cramer's transactional
interpretation; and (iv) the Hess--Philipp
analysis of extended parameter spaces.
\end{abstract}
\section{Introduction}
In one of his thought-provoking discussions
of the two-slit experiment, Feynman~\cite{feynman}
expressed the view that ``it is safe
to say that no one understands quantum mechanics.
\mbox{[\ldots]} Nobody knows how it can be like
that.''  Yet 80 years of research in the foundations
of the theory have led a growing
number of investigators not to share Feynman's
fatalism.  They, in fact, have turned his assessment
into a challenge, by asking ``How can
the world be for quantum mechanics to be true?''
I have argued elsewhere~\cite{marchildon}, following
others~\cite{fraassen}, that interpreting
the theory consists in providing a precise answer
to this question.  Moreover, I believe that
providing more than one possible and consistent
answer, far from introducing confusion, brings
instead additional understanding, and may even
stimulate the imagination.

Long-distance quantum correlations, first pointed
out by Einstein, Podolsky and Rosen (EPR)~\cite{einstein},
and sharply investigated by Bell~\cite{bell},
have long been considered paradoxical and in need
of explanation.  It is the purpose of this
contribution to briefly review and analyze
four different approaches
through which one can make sense of them.

\section{Long-distance correlations}
\begin{figure}[htb]
\begin{center}
\epsfig{file=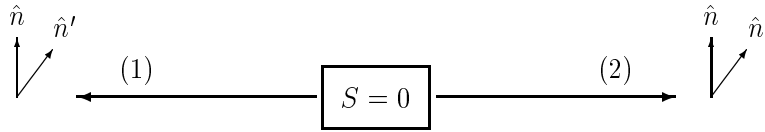,width=4in}
\caption{Two particles prepared in the singlet state
and leaving in opposite directions.}
\label{epr}
\end{center}
\end{figure}

Following Bohm~\cite{bohm1}, I consider two
spin 1/2 particles prepared in the singlet
state $|\chi\rangle$
and leaving in opposite directions (Fig.~\ref{epr}).
On each side, an apparatus can measure the
component of the spin of the associated particle
either along axis $\hat{n}$ or along axis~$\hat{n} '$. 

In a nutshell, the paradoxical features of the
arrangement can be expressed as follows:
\begin{enumerate}
\item Quantum mechanics predicts, and experiments
confirm, that there is perfect anticorrelation
when the spins of both particles are measured
along the same axis.
\item This seems to suggest, against conventional
quantum mechanics, that all spin components have values
even before they are measured.  This assertion
is an instance of \emph{local realism}.
\item Quantum mechanics predicts, and experiments
confirm, that the spin correlations are in general
given by
\begin{equation}
\langle \chi | \vec{\sigma}_{1} \cdot \hat{n}
\otimes \vec{\sigma}_{2} \cdot \hat{n}' | \chi \rangle
= - \hat{n} \cdot \hat{n}' .
\label{corr}
\end{equation}
\item Conventional wisdom holds that local
realistic theories imply, against quantum
mechanics, that the spin correlations
satisfy the Bell inequalities.  These are
inconsistent with Eq.~\ref{corr} and are
experimentally violated.
\end{enumerate}

\section{Copenhagen and related views}
In his reply to the EPR paper,
Bohr~\cite{bohr} emphasized the holistic
aspect of measurement.  For him, the whole experimental
setup is inseparable.  The measurement of a physical
quantity on one side fundamentally alters the
conditions of the measurement of a conjugate
variable on the other side.  It prevents the
definition of meaningful ``elements of reality''
pertaining to one part of a system only.
Since no value can be ascribed to an observable
outside its measurement context, the inference from
statement~(1) to statement~(2) in the last section is,
for Bohr, unwarranted.

Among the leading exponents of the Copenhagen
interpretation, Heisenberg~\cite{heisenberg}
has expressed the view that the state vector
represents knowledge rather than the
state of an independent object.
The development of quantum information theory
has led to renewed interest in this \emph{epistemic
view} of quantum states.  To quote a recent
column~\cite{fuchs}, ``the time dependence of the
wave function does not represent the evolution
of a physical system.  It only gives the
evolution of our probabilities for the
outcomes of potential experiments on
that system.''  In the epistemic view, or at least
in the more radical variants of it,
there are no microscopic
carriers of elements of reality.  The state vector
is simply a device to predict correlations
between distant local measurements.  The correlations
don't stand in need of further explanation.

That quantum mechanics is about information is
also stressed in the relational view advocated by
Rovelli~\cite{rovelli1,rovelli2}.  In relational
quantum mechanics, all systems (including apparatus)
are quantum mechanical.  An observable can have a
value with respect to an observer and not with respect
to another.

Specifically, spin $\vec{\sigma}_2 \cdot \hat{n} '$
has a value for observer $O_1$ only if $O_1$ measures it,
or measures the result obtained by $O_2$.
The meaningful correlations are not those between
$\vec{\sigma}_1 \cdot \hat{n}$ measured by $O_1$ and
$\vec{\sigma}_2 \cdot \hat{n} '$ measured by $O_2$,
but (say) those between $\vec{\sigma}_1 \cdot \hat{n}$ and
$\vec{\sigma}_2 \cdot \hat{n} '$ both measured by $O_1$.
Hence there is no problem with locality.  The price
one has to pay for this resolution of the paradox
is, however, a rather significant weakening of realism.
Specifically, statements like ``the measurement
of this observable has yielded that value'' no longer
hold in an absolute way.

\section{Bohmian mechanics}
The Schrödinger wave function for two spinless
particles can be written as
\begin{equation}
\Psi (\vec{r}_1, \vec{r}_2, t) 
= \rho (\vec{r}_1, \vec{r}_2, t)
\exp\left\{ \frac{i}{\hbar} S(\vec{r}_1, \vec{r}_2, t)
\right\} .
\end{equation}
In Bohmian mechanics, the particles follow deterministic
trajectories governed by~\cite{bohm2,holland}
\begin{equation}
\vec{v}_1 = \frac{1}{m_1} \vec{\nabla}_1 S , \qquad
\vec{v}_2 = \frac{1}{m_2} \vec{\nabla}_2 S .
\label{traj}
\end{equation}
The statistical predictions of quantum mechanics
are recovered by postulating that the particles
are drawn from an ensemble with probability
density $|\Psi|^2 = \rho^2$.

For a factorizable wave function like
\begin{equation}
\Psi (\vec{r}_1, \vec{r}_2, t)
= \psi_1 (\vec{r}_1, t) \psi_2 (\vec{r}_2, t) ,
\end{equation}
we get $S(\vec{r}_1, \vec{r}_2, t) = S_1(\vec{r}_1, t)
+ S_2(\vec{r}_2, t)$, and the motion of particle 1 is
independent of what happens to particle 2. 
But for an entangled wave function like
\begin{equation}
\Psi (\vec{r}_1, \vec{r}_2, t)
= \sum_k \psi_1^{(k)} (\vec{r}_1, t) \psi_2^{(k)}
(\vec{r}_2, t) , 
\end{equation}
the function $S(\vec{r}_1, \vec{r}_2, t)$ does not
break up into the sum of a function of $\vec{r}_1$
and a function of $\vec{r}_2$.  What happens to particle~2
instantaneously affects the motion of particle~1.
From this one may be tempted to conclude that
Bohmian mechanics will allow for superluminal transfer
of information.  This is indeed the case if state
preparation is not suitably restricted~\cite{valentini}.
But if particles are prepared with a probability
density $|\Psi(\vec{r}_1, \vec{r}_2, t_0) |^2$ at
time $t_0$, they evolve into a density
$|\Psi(\vec{r}_1, \vec{r}_2, t) |^2$ at any time $t$,
and one can show that no superluminal transfer of
information is possible.

To incorporate spin in Bohmian mechanics, one adds
spinor indices to the wave function, in such a way
that $\Psi \rightarrow \Psi_{i_1 i_2}$.  There can
be several ways to associate particle spin vectors
with the wave function~\cite{bohm2}, but one way
or other they involve the expressions
\begin{equation}
\vec{s}_1 = \frac{\hbar}{2 \Psi^{\dagger} \Psi}
\Psi^{\dagger} \vec{\sigma}_1 \Psi , \qquad
\vec{s}_2 = \frac{\hbar}{2 \Psi^{\dagger} \Psi}
\Psi^{\dagger} \vec{\sigma}_2 \Psi .
\end{equation}

In the singlet state, the initial wave function
typically has the form
\begin{equation}
\Psi = \psi_1(\vec{r}_1) \psi_2(\vec{r}_2)
\frac{1}{\sqrt{2}} (u_{1+} u_{2-} - u_{1-} u_{2+}) ,
\end{equation}
in obvious notation.  With such wave function,
it is easy to show that $\vec{s}_1 = 0$ and
$\vec{s}_2 = 0$.  That is, both particles
initially have spin zero.  This underscores
the fact that in Bohmian mechanics, values of
observables outside a measurement context do not
in general coincide with operator eigenvalues.

Spin measurement was analyzed in detail in
Refs.~\cite{dewdney1} and~\cite{dewdney2}.
In the EPR context, in particular, Dewdney,
Holland and Kyprianidis first wrote down the
two-particle Pauli equation adapted to the situation
shown in Fig.~\ref{epr}.  With Gaussian initial
wave packets $\psi_1$ and $\psi_2$, the equation can
be solved under suitable approximations.
Bohmian trajectories can then
be obtained by solving Eq.~\ref{traj}.  These involve
the various components of the two-particle wave
function in a rather complicated way, and must be
treated numerically.

Suppose that the magnetic field in the Stern--Gerlach
apparatus on the left of Fig.~\ref{epr} is oriented in
the $\hat{n}$ direction.  Consider the case where
particle~1 enters that apparatus much before particle~2
enters the one on the right-hand side.  What was
shown was the following.  When particle~1 enters the
apparatus along a specific Bohmian trajectory, the
various forces implicit in Eq.~\ref{traj} affect both
the trajectory and the spin vector, the latter building
up through interaction with the magnetic field.
The beam in which particle~1 eventually ends up depends
on its initial position.  If particle~1
ends up in the upper beam of the Stern--Gerlach
apparatus, its spin becomes aligned
with $\hat{n}$.  Meanwhile there is an
instantaneous action on particle~2,
simultaneously aligning its spin in the
$-\hat{n}$ direction.  Similarly, if particle~1's
initial position is such that it ends
up in the lower beam, its spin becomes aligned with
$-\hat{n}$, and the spin of particle~2 simultaneously
aligns in the $\hat{n}$ direction.

Thus the nonlocal forces inherent in Bohmian
mechanics have, once the measurement of the spin
of particle~1 has been completed, resulted in particle~2
having a spin exactly opposed.  It is then easy
to see that if particle~2 later enters a 
Stern--Gerlach apparatus with magnetic field
oriented in the $\hat{n} '$ direction, its deflection
in the upper or lower beam will precisely reproduce
the correlations of Eq.~\ref{corr}.

\section{The transactional interpretation}
Cramer's transactional interpretation~\cite{cramer1,cramer2}
is inspired by the Wheeler--Feynman electromagnetic
theory, in which advanced electromagnetic waves are
as important as retarded waves.

In this interpretation,
a quantum process (e.g. the emission of an $\alpha$
particle, followed by its absorption by one of several
detectors) is held to involve the exchange of offer
waves (solutions of the Schr\"{o}dinger equation) and
confirmation waves (complex conjugates of the former).
The confirmation waves propagate backward in time.

Suppose that D, at point $\vec{r}$, is one of
a number of detectors that can absorb the particle.
The offer wave, emitted at $t_0$ from the $\alpha$
particle source, will arrive at~D with an
amplitude proportional to $\psi(\vec{r}, t)$,
the Schr\"{o}dinger wave function.  The confirmation
wave produced by~D is stimulated by the offer
wave, and Cramer argues that it arrives back at
the source with an amplitude proportional to
$\psi(\vec{r}, t) \psi^*(\vec{r}, t)$
= $|\psi(\vec{r}, t)|^2$.  Similar offer and
confirmation waves are exchanged between the
source and all potential detectors, and all
confirmation waves reach the source exactly
at~$t_0$, the time of emission.  Eventually,
what Cramer calls a \emph{transaction} is
established between the source and one of the
detectors, with a probability proportional to
the amplitude of the associated confirmation
wave at the source.  The quantum process is
then completed.

\begin{figure}[htb]
\begin{center}
\epsfig{file=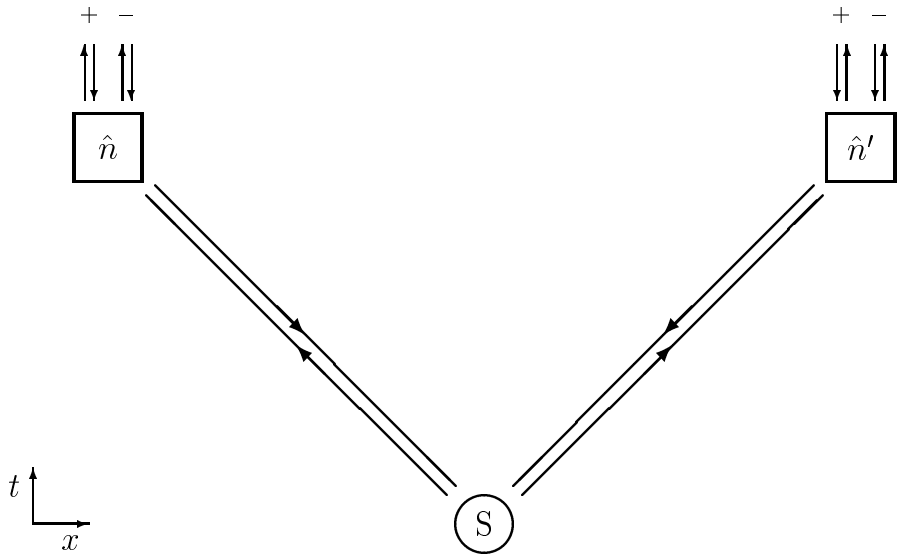,width=4.7in}
\caption{Offer waves (upward arrows) and confirmation
waves (downward arrows) in the EPR setup.}
\label{transac}
\end{center}
\end{figure}

Fig.~\ref{transac} is a space-time representation
of an EPR setup, in the transactional
interpretation.  Arrows pointing in the positive
time direction label offer waves, and those pointing
in the negative direction label confirmation waves.
Two particles are emitted by the source, and in Cramer's
sense each particle can be absorbed by two detectors,
corresponding to the two beams in which each particle
can emerge upon leaving its Stern--Gerlach apparatus.

Let us focus on what happens on the left-hand side.
An offer wave is emitted by the source, and in going
through the Stern--Gerlach apparatus it splits into
two parts.  One part goes into the detector labelled~$+$,
and the other goes into detector~$-$.  Each detector
sends back a confirmation wave, propagating backward
in time through the apparatus and reaching
the source at the time of emission.  A transaction
is eventually established, resulting in one of the
detectors registering the particle.  A similar
process occurs on the right-hand side, with one of
the two detectors on that side eventually
registering the associated particle.

If offer and confirmation waves represent causal
influences of some sort, one can see that these
influences can be transmitted between the
spacelike-separated detectors on different sides
along paths that are entirely timelike or lightlike.
In this way, the EPR correlations are explained
without introducing any kind of superluminal motion.

\section{Extended parameter space}
Bohmian mechanics and Cramer's transactional
interpretation explain the long-distance
quantum correlations by means of channels which,
although not allowing for the superluminal transfer
of classical bits of information, involve
causal links of some sort between
spacelike-separated instruments.  In recent
work, Hess and Philipp~\cite{hess1} have argued
that the correlations might be understood without
appealing to such links.

In the original proof of his inequality,
Bell~\cite{bell} assumed that the state of the
particle pair is characterized by a hidden
variable $\lambda$, which represents one of the
values of a random variable $\Lambda$.  He
further assumed that the result of
measuring (say) the $\hat{n}$ component of the
spin of particle~1 is fully determined by $\lambda$
and $\hat{n}$.  He supposed, however, that the
result of measuring the $\hat{n}$ component of the
spin of particle~1 does not depend on which
component of the spin of particle~2 is measured.
The latter assumption embodies the prohibition
of superluminal causal influences, and is fully
endorsed by Hess and Philipp.

Hess and Philipp point out that Bell's proof,
as well as all subsequent proofs of similar
inequalities, make use of parameter spaces
that are severely restricted.  They introduce
much more general spaces.  Like Bell, they assume
that pairs of particles emitted in the singlet
state are characterized by a random variable
$\Lambda$, which is stochastically independent
on the settings on both sides.
But then they associate with each
measuring instrument random variables
$\Lambda^{(1)}_{\hat{n}} (t)$ and
$\Lambda^{(2)}_{\hat{n} '} (t)$, which depend
both on the setting of the instrument and on the
time.  The result of measuring, say, the $\hat{n}$
component of the spin of particle~1 at time $t$,
is taken to be a deterministic function of
$\Lambda$ and $\Lambda^{(1)}_{\hat{n}} (t)$.

Several important remarks should be made at this
stage.  Firstly, and in the spirit of standard quantum
mechanics, neither particle has a precise value
of any of its spin components before measurement.
Rather, the particles and the instruments
jointly possess information
that is sufficient for deterministic values to obtain
upon measurement.  Secondly, the dependence of
the instruments' random variables on some universal
time allows for a stochastic dependence of
measurement results on one another, conditional on
$\Lambda$, if the measurements are performed at
correlated times in the two wings.
And yet thirdly, the measurement result on one side
can be stochastically independent on the setting
on the other side.

With such extended parameter spaces, Hess and
Philipp have shown that the standard proofs of
the Bell inequalities come to a halt.  Such
proofs typically assume that the two particles,
once they have left the source, simultaneously
have well-defined values of more than one spin
component.  But in the extended parameter space
approach, spin components get values only upon
measurement.  Counterfactual reasoning is allowed
only in the sense that had a different spin
component been measured, it would have yielded a
definite and deterministic value.  But that value
does not exist before measurement.  And since the
measurement of different spin components requires
incompatible apparatus, different spin components
of the same particle cannot have values at
the same time. But spin components of both particles
measured at correlated times in the two wings
can be stochastically dependent, through the
dependence of the instrument random variables on time.

In experimental tests of the Bell inequalities,
spin measurements on a given pair were performed
in a time frame many orders of magnitude smaller
than the time interval between successive
measurements on two different pairs.  It is
therefore conceivable that a time dependence
of the instrument random variables, having no
effect on such properties as perfect anticorrelation
for particles in the same run, could reproduce the
quantum-mechanical long-distance correlations
observed on runs performed at different times.
Such runs would not sample the quantities that
appear in the standard forms of the Bell
inequalities.

Hess and Philipp also proposed an explicit
model of Einstein-local random variables that lead
to violations of the
Greenberger--Horne--Zeilinger equations~\cite{hess2},
violations that experiments claimed to have
observed~\cite{pan}.

\section{Summary and conclusion}
The long-distance quantum correlations
and the violation of Bell inequalities can
be understood in a number a different ways,
four of which were reviewed here.

In the Copenhagen and epistemic views,
correlations are basically dealt with by
relaxing the requirements of explanation.
In Bohmian mechanics, instantaneous
interactions orient the spin of the second
particle while the spin of the first one is
measured, but restrictions on state preparation
prevent the superluminal transfer of information.
In the transactional interpretation, advanced waves
provide for a communication channel between
spacelike-separated detectors.  In the
Hess--Philipp approach, finally, correlations
are explained through instrument random variables
that depend both on setting and on time.

I have attempted to illustrate the idea
that a theory is made clearer through the
display of various models that make it true.
This is the process of interpretation, and
in connection with it one should be wary of
identifying consequences of the formalism of
quantum mechanics with consequences of specific
interpretations of it.  This, unfortunately,
has not always been done, as the example of
Ref.~\cite{zeilinger} still shows.

\section*{Acknowledgments}
This work was supported by the Natural Sciences
and Engineering Research Council of Canada.
I thank Karl Hess for comments on the manuscript.
I am grateful to many participants in the workshop
for discussions, and especially to Walter Philipp,
who has sadly left us since and will be keenly
remembered.

\end{document}